\documentclass[12pt,onecolumn,draftclsnofoot]{IEEEtran}
\IEEEoverridecommandlockouts

\usepackage{cite}
\usepackage{amsmath,amssymb,amsfonts}
\usepackage{algorithmic}
\usepackage{graphicx}
\usepackage{textcomp}
\usepackage{xcolor}
\usepackage{braket}
\usepackage{lineno}

\def\BibTeX{{\rm B\kern-.05em{\sc i\kern-.025em b}\kern-.08em
    T\kern-.1667em\lower.7ex\hbox{E}\kern-.125emX}}
\begin{document}

\title{Sub-nanosecond control for spin-defect quantum memories with a low-cost, compact FPGA platform}

\author{
\IEEEauthorblockN{
Victor Marcenac\textsuperscript{1,*},
Tommy Nguyen\textsuperscript{2,*},
Julie Chen\textsuperscript{1},
Weitao He\textsuperscript{2},
Enrique Garcia\textsuperscript{1},
Yuyang Han\textsuperscript{2},\\
Bethany E. Matthews\textsuperscript{3},
Tiamike Dudley\textsuperscript{4,5},
Andrew Mounce\textsuperscript{5},
Kai-Mei C. Fu\textsuperscript{1,2,3},
Maxwell F. Parsons\textsuperscript{1,2,\textdagger}
}
\and
\IEEEauthorblockA{
\textsuperscript{1}Department of Electrical \& Computer Engineering, University of Washington, Seattle, Washington 98195, USA\\
\textsuperscript{2}Department of Physics, University of Washington, Seattle, Washington 98195, USA\\
\textsuperscript{3}Pacific Northwest National Laboratory, Richland, Washington 99352, USA\\
\textsuperscript{4}Electrical and Computer Engineering Department, The University of New Mexico, Albuquerque, New Mexico 87106, USA\\
\textsuperscript{5}Center for Integrated Nanotechnologies, Sandia National Laboratories, Albuquerque, New Mexico 87123, USA\\
\textsuperscript{*}These authors contributed equally.\\
\textsuperscript{\textdagger}Corresponding author: mfpars@uw.edu
}
}

\maketitle

\begin{abstract}
Dynamical decoupling techniques are widely used to characterize and control the environments of solid-state quantum defects, enabling solid-state quantum memories and nanoscale quantum sensors. However, resolution is often limited by the timing granularity of control hardware, which can undersample narrow spectral features and distort extracted parameters. Here, we demonstrate sub-nanosecond timing control on an inexpensive FPGA-based platform by extending the open-source QICK (Quantum Instrumentation Control Kit) framework using a waveform-offset method. This approach achieves an effective timing resolution of 200~ps on an RF system-on-chip device without modification to the underlying hardware.  We apply this capability to dynamical decoupling spectroscopy of nitrogen-vacancy centers in diamond, enabling precise extraction of hyperfine couplings of individual $^{13}\mathrm{C}$ nuclear spins and resolving spectral features that are otherwise undersampled.  These results demonstrate that high-resolution, device-level characterization of spin-based quantum memories can be achieved using flexible, inexpensive control hardware, providing a scalable alternative to commercial arbitrary waveform generators.
\end{abstract}


\section{Introduction}
Quantum memories are a critical resource for quantum information processing, enabling long-lived storage of quantum states and providing scalable entanglement distribution in quantum networks \cite{DuanEtAl2001, HeshamiEtAl2016, AzumaEtAl2023}. Solid-state spin defects in wide band gap materials, such as the nitrogen-vacancy (NV) center in diamond, are among the most promising platforms for realizing such memories due to their optical interface, long electronic spin coherence, and the presence of proximal nuclear spins that can serve as high-fidelity, long-lived quantum registers \cite{TaminiauEtAl2012}.

Two key challenges limit the scalability of color center quantum memories. First, the size of the usable nuclear spin register is fundamentally constrained by the ability to identify, resolve, and coherently control individual spins within a complex spin environment. Second, each device exhibits a unique local nuclear configuration, requiring detailed characterization to identify usable qubits and extract their coupling parameters. Recent work suggests that detailed knowledge of multi-qubit entanglement dynamics can enable hardware-efficient, multi-qubit gate protocols tailored to a given nuclear spin environment~\cite{TakouPreciseControlEntanglement2023, gnassoQuantifyingElectronnuclearSpin2026, MinnellaEtAl2026}. However, such approaches are inherently device-specific and rely on precise knowledge of the full spin Hamiltonian, further increasing the need for high-resolution characterization. As systems scale toward larger memories and networked architectures, this characterization overhead becomes a critical bottleneck.

A powerful tool for probing the nuclear spin environment is dynamical decoupling spectroscopy, such as Carr-Purcell-Meiboom-Gill (CPMG) sequences, which transfers dynamics of the local magnetic field at the defect into measurable coherence modulations \cite{ZhaoEtAl2012, BiercukEtAl2011, Hernandez-Gomez2018}.  When those field dynamics are due to surrounding spinful nuclei, the CPMG signal can be used to map the nuclear spin environment \cite{TaminiauEtAl2012}. However, extracting quantitative information from these measurements places stringent demands on control hardware. In particular, resolving weak hyperfine couplings and narrow spectral features requires fine control of inter-pulse timing, often beyond the native resolution of conventional arbitrary waveform generators (AWGs), which typically operate at nanosecond-scale granularity. This limitation can obscure spectral features, reduce fitting accuracy, and ultimately restrict the available register size.

In this work, we demonstrate an accessible method for achieving sub-nanosecond timing control in spin-defect experiments using an FPGA-based platform. This improved timing resolution enables high-fidelity dynamical decoupling spectroscopy, allowing extraction of hyperfine coupling parameters of individual $^{13}\mathrm{C}$ nuclear spins with sub-kHz precision. In addition, the enhanced resolution reveals spectral features that are otherwise undersampled, including peaks that cannot be reconciled with the expected $^{13}\mathrm{C}$ Larmor frequency and hyperfine model under the applied magnetic field.

To realize this capability, we develop a high-resolution control approach based on an extension of the open-source Quantum Instrumentation Control Kit (QICK) framework \cite{DingEtAl2024}, building on the QICK-DAWG variant for color centers \cite{RiendeauEtAl2023}. By implementing a waveform-offset method on an RF system-on-chip (RFSoC) platform, we achieve sub-nanosecond timing control with an effective resolution of 200~ps without modification to the underlying hardware, providing a scalable and low-cost alternative to high-end AWG-based systems.

\section{FPGA Control System}

\subsection{RFSoC Control Architecture}
Quantum control of spin qubits requires precise timing of microwave pulses synchronized with optical initialization and readout. Even basic pulse protocols, such as Rabi oscillations, rely on accurate pulse timing for maximizing the signal to noise ratio (SNR). More advanced protocols, including dynamical decoupling sequences such as CPMG, require precise control of inter-pulse delays. Because these sequences map spin interactions into narrow features in the delay parameter $\tau$, insufficient timing resolution leads to undersampling and distorted spectral features.

Traditional control systems for spin-defect experiments rely on multiple synchronized instruments, which are complex, expensive, and difficult to scale.  FPGA-based platforms such as ARTIQ \cite{kasprowiczARTIQSinaraOpen2020} and QICK \cite{DingEtAl2024} provide a compelling alternative by enabling deterministic timing, tight integration of control and readout, and flexible programmability within a single framework.  The ``Defect Arbitrary Waveform Generator'' variant of QICK (QICK-DAWG) extends this framework to optical spin systems by incorporating microwave control, photon counting, and optical synchronization with a smaller hardware footprint \cite{RiendeauEtAl2023}.

The RFSoC 4x2 platform used by QICK-DAWG and in this work contains a 9.85~GHz Digital‑to‑Analog Converter (DAC). The QICK-DAWG firmware supports a 4.915~GHz DAC operation and a 307.2~MHz FPGA sequencer. This architecture introduces a fundamental mismatch in timing resolution: while waveform samples are defined at the DAC rate ($\sim$200~ps), pulse scheduling is constrained to the FPGA sequencer clock, resulting in a native timing granularity of $\sim$3.25~ns. As a result, pulse timing in QICK-DAWG is restricted to integer multiples of the sequencer clock cycle, even though the underlying waveform representation supports much finer timing resolution. This limitation is not specific to QICK, but is a general feature of FPGA-based control systems that separate instruction timing from waveform playback.  This mismatch between available waveform resolution and achievable pulse timing motivates the development of methods to access sub-clock-cycle timing control within the constraints of the FPGA architecture.

\begin{figure}
    \centering
    \includegraphics[width=0.5\columnwidth]{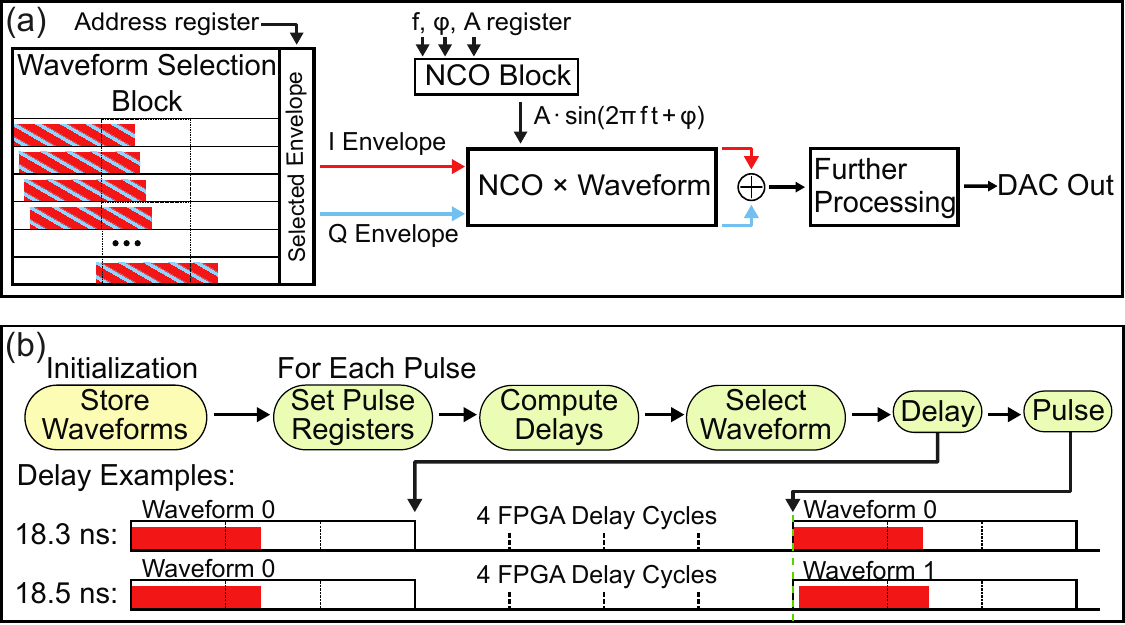}
    \caption{(a) Conceptual diagram of the microwave (MW) DAC pipeline. In-phase (I) and quadrature (Q) envelopes are mixed with a sinusoidal carrier generated by a numerically-controlled oscillator (NCO). The NCO sets the carrier frequency $f$, phase $\phi$, and amplitude $A$, producing a sinusoidal output in time $t$. The mixed I/Q signals are then combined and processed through additional filtering and correction steps to produce the final MW output. The staggered waveform bank is shown conceptually, with individual waveforms offset in $\sim$200~ps (single-sample) steps. (b) Illustration of the procedure for programming an arbitrary delay between two pulses. During initialization, the staggered waveforms are loaded into memory. At runtime, for each pulse the controller sets the pulse parameters (frequency, amplitude, phase), computes coarse and fine delays (including dead time from the previous pulse), updates the waveform-selection registers, applies the programmed delay, and triggers the pulse.}
    \label{fig:1}
\end{figure}

\subsection{Sub-Nanosecond Timing via Waveform Offsets}

To bridge this mismatch, we implement a coarse–fine timing decomposition that enables effective sub-nanosecond control without modifying the underlying hardware, summarized in Fig. \ref{fig:1}

In this approach, a desired pulse delay is expressed as the sum of: a coarse delay, specified in units of the sequencer clock cycle, and a fine delay, specified in units of the DAC sample period.

The coarse delay determines when a pulse is scheduled by the sequencer, while the fine delay is implemented by selecting from a precomputed bank of waveform segments. Each waveform in the bank represents an identical pulse shape shifted in time by an integer number of DAC samples. By choosing the appropriate waveform index at runtime, the pulse can be aligned to the DAC sample grid with sub-nanosecond precision.

Operationally, the desired delay is specified in units of DAC samples and decomposed into coarse and fine components via integer division and modulo operations. The coarse component sets the sequencer delay, while the remainder selects the waveform index. Because the upsampling factor between the sequencer and DAC clocks is a power of two, this decomposition can be implemented efficiently using bitwise operations.

The waveform bank has finite memory ($2^{16}$ samples per DAC channel) imposing a trade-off between the achievable timing resolution and the total number and duration of distinct pulse segments. The coarse–fine decomposition also introduces per-pulse instruction overhead, limiting the minimum spacing between successive waveforms in long pulse sequences.

Beyond these considerations, the RFSoC provides two complementary mechanisms for phase control. The numerically controlled oscillator (NCO) sets the carrier phase with 32-bit resolution and can be updated at the sequencer rate without additional memory overhead, enabling efficient and highly precise phase shifts between pulses. For sequences requiring rapid or intra-pulse phase changes, phase can instead be encoded directly in the I/Q waveform envelopes, which are defined with 16-bit amplitude resolution per quadrature. Together, these approaches enable phase-coherent control while efficiently managing waveform memory.

\subsection{Application to CPMG Spectroscopy}
We apply the coarse–fine timing method to implement the Carr–Purcell–Meiboom–Gill (CPMG) sequence, a widely used dynamical decoupling protocol for spin bath spectroscopy \cite{TaminiauEtAl2012}. The sequence consists of repeated $\tau - \pi - \tau$ segments, where precise control of the inter-pulse delay $2\tau$ determines the spectral signature (Fig. \ref{fig:3}a). Using the phase control available with QICK-DAWG, we symmetrize the decoupling sequence (CPMGXY) by alternating the phase of the $\pi$ pulses around the X and Y axes (X-Y-X-Y-Y-X-Y-X) to mitigate pulse errors. 

Without implementing our waveform sequencing method, $\tau$ is restricted to integer multiples of the sequencer clock cycle ($\sim$3.25~ns), resulting in a discretized sampling of the filter function. This limits the ability to resolve narrow spectral features and accurately extract hyperfine couplings from closely spaced resonances.

Using the coarse–fine timing method, $\tau$ is defined and swept in units of DAC samples ($\sim$200~ps), enabling significantly finer sampling of the CPMG filter function. Both $\pi$ and $\pi/2$ pulses are implemented as waveform banks of sample-shifted replicas, allowing each pulse to be placed with sub-nanosecond precision.

For each pulse, the desired start time—defined relative to the end of the previous pulse—is specified in DAC sample units and decomposed using the fixed 16:1 ratio between the DAC and sequencer clocks. Integer division determines the sequencer delay, while the remainder selects the waveform offset, providing deterministic alignment to the DAC sample grid. As in the general method, this decomposition is implemented efficiently using bitwise operations.

This approach enables dense and uniform sampling of the CPMG response, improving the ability to resolve weakly coupled nuclear spins and to precisely extract hyperfine parameters.

\section{Nuclear spin bath characterization}

\subsection{NV Centers as a representative system}

While we focus here on the nitrogen-vacancy (NV) center in diamond as a well-understood and extensively characterized platform, the techniques demonstrated in this work are broadly applicable to a wider class of optically addressable spin defects. Several emerging color centers exhibit similar electronic–nuclear spin structures that support multi-qubit quantum registers, including silicon-vacancy (SiV) \cite{knautEntanglementNanophotonicQuantum2024}, germanium-vacancy (GeV) \cite{gundlapalliHighFidelitySingleShotReadout2025}, and tin-vacancy (SnV) centers in diamond \cite{beukersControlSolidStateNuclear2025}, as well as defect systems in silicon carbide \cite{tissotNuclearSpinQuantum2022}. In addition, T-centers in silicon have recently attracted interest as a telecom-compatible spin-photon interface with long-lived nuclear spin memories \cite{songEntanglementNuclearSpin2026}. These platforms share the key feature of a central electronic spin coupled to a local nuclear environment, where precise characterization of hyperfine interactions is essential for identifying and controlling quantum registers. In this context, the NV center serves as a representative test system for developing and validating high-resolution control and characterization techniques that can be extended to these emerging material platforms.

\begin{figure}
    \centering
    \includegraphics[width=0.5\columnwidth]{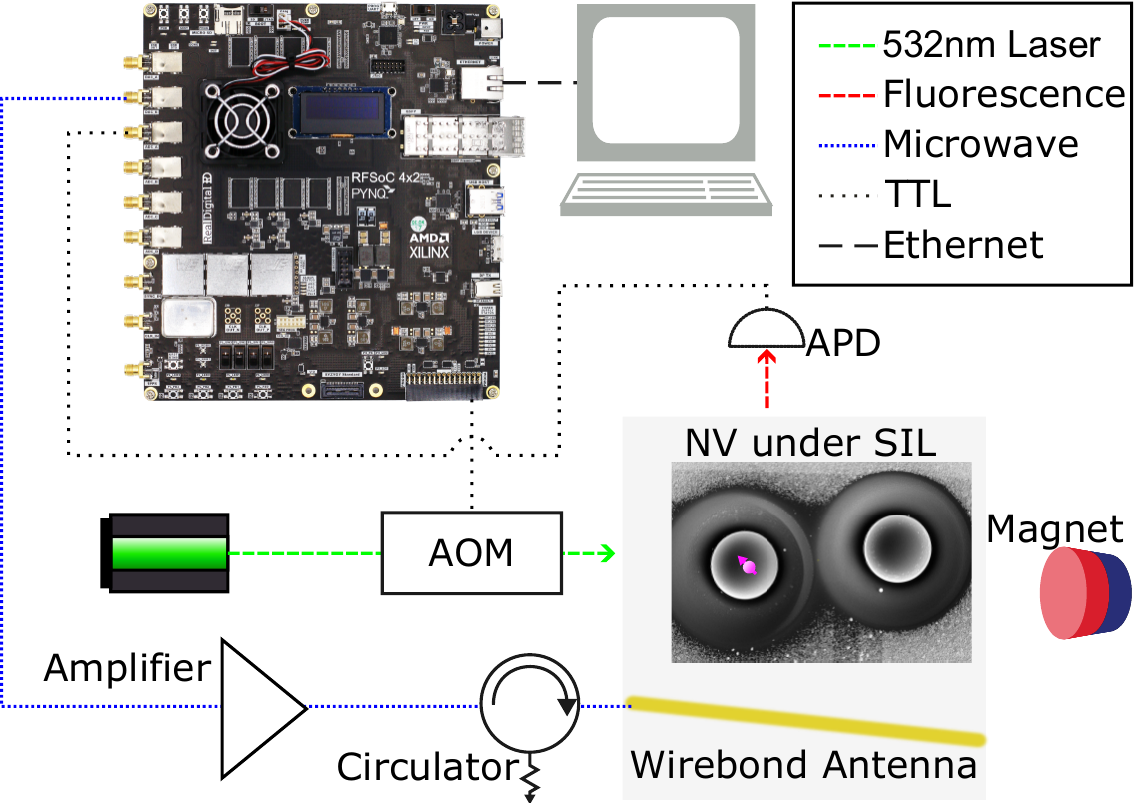}
    \caption{Control schematic of the room‑temperature NV experimental setup. The RFSoC~4$\times$2 generates microwave (MW) control pulses, gates the green‑laser excitation via a PMOD‑driven TTL to the AOM driver, and records photon‑count TTLs from an avalanche photodiode. MW pulses from the DAC are amplified and routed through a circulator to the wirebond antenna positioned $\sim$25~\textmu m above the diamond and $\sim$15~\textmu m lateral to the SIL. A host computer issues high‑level QICK‑DAWG pulse programs to the RFSoC via a remote-proxy object.}
    \label{fig:2}
\end{figure}

\subsection{Room-temperature characterization in a user-facility confocal microscope}

We employ the QICK-DAWG platform as a unified control system for synchronized optical excitation, microwave manipulation, and photon counting within a room-temperature confocal microscope (control schematic shown in Fig. \ref{fig:2}). This integration enables all calibration and measurement routines to be executed within a single control framework, providing fine timing resolution while minimizing footprint.  The compact hardware configuration is especially advantageous in a shared user-facility confocal microscope, where RF equipment must be swapped out between measurements to make the instrument available to other users.

The NV sample we use for room-temperature characterization has a natural $^{13}\mathrm{C}$ abundance of nuclear spins (1.1\%). The electronic spin is initialized into the $m_s = 0$ ground state through optical excitation with a 532~nm laser, and readout is performed through spin-dependent fluorescence. Photon collection efficiency is enhanced by a solid immersion lens (SIL). A static magnetic field of approximately $\sim365$~G is applied using a neodymium permanent magnet and aligned to the NV symmetry axis by maximizing the photoluminescence while minimizing the optically detected magnetic resonance derived misalignment angle \cite{liExcitedstateDynamicsOptically2024} to lift the $m_s=\pm1$ degeneracy. Microwaves (MW) are synthesized by the RFSoC with our QICK-DAWG implementation and sent to the sample through a wirebonded antenna to drive spin transitions between the electronic $\ket{m_s=0}$ and $\ket{m_s=-1}$ states which define our qubit.  At this field, we observe a single dominant hyperfine resonance at $1.845$~GHz, indicating a high degree of $^{14}$N nuclear spin polarization, consistent with operation near the ESLAC around $500$ G \cite{BusaiteEtAl2020}. We use a MW power corresponding to a Rabi frequency of $9.18$~MHz to drive electron spin flips.

At the saturation intensity of the green excitation light, we measure a longitudinal relaxation time of $T_1 = 0.90 \pm 0.10$~ms. We attribute this limitation in $T_1$ to laser acousto-optic modulator (AOM) leakage. By reducing the laser power, we extend the relaxation time to $T_1 = 1.71 \pm 0.11 $~ms, which provides a reasonable compromise between coherence and SNR. We also measure a Hahn echo coherence time of $T_2 = 0.68 \pm 0.22$~ms. These values confirm that the NV center is suitable for nuclear spin spectroscopy CPMG experiments requiring extended coherence times. 

\begin{figure*}
    \centering
    \includegraphics[width=1\textwidth]{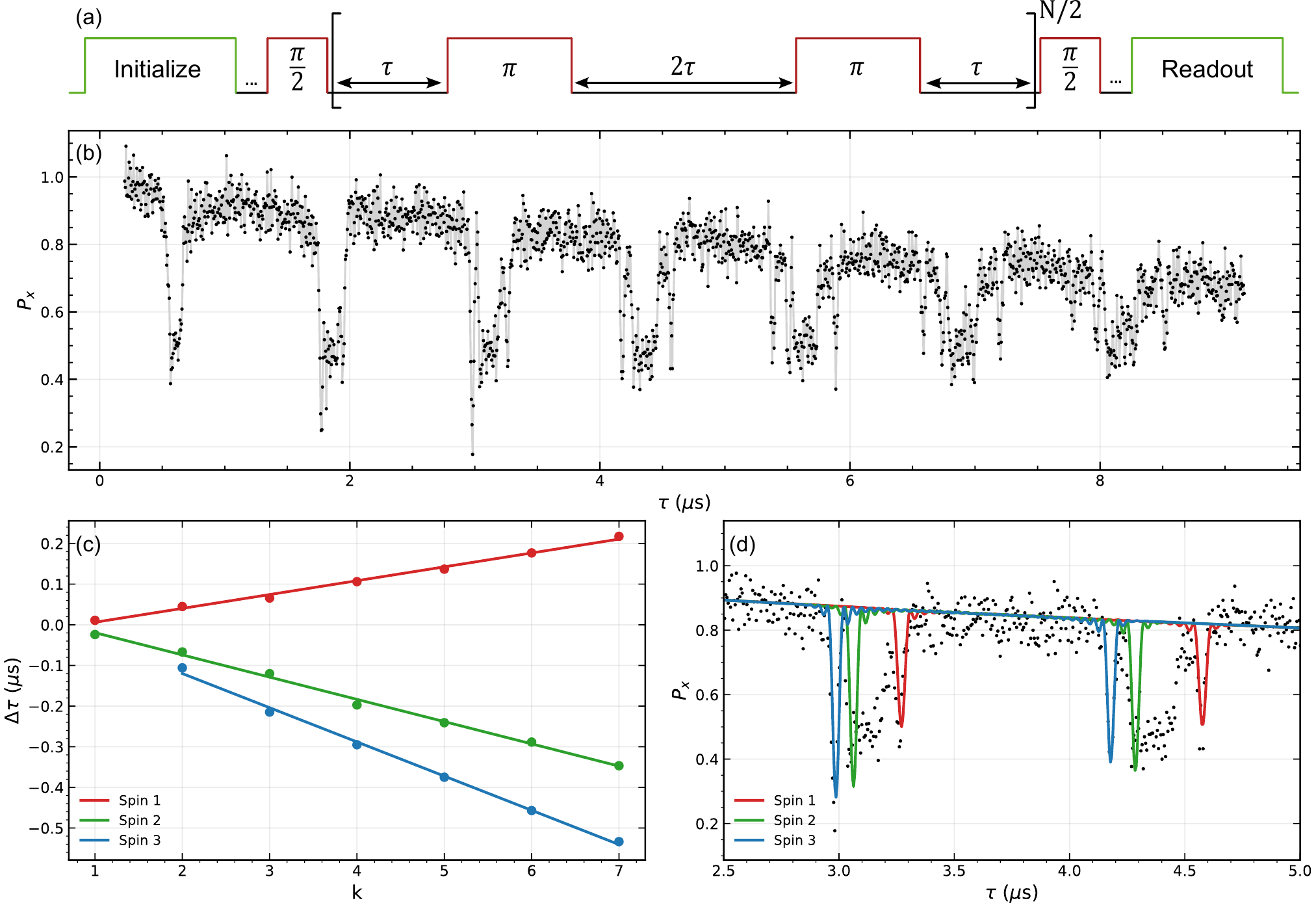}
    \caption{Characterization of the nuclear spin environment (a) Pulse sequence controlled by QICK-DAWG. A 532~nm green laser is used to initialize the electron into the $NV^-$ charge and $\ket{m_s}=0$ spin state. The electron spin is prepared in a superposition state with a starting $\pi/2$ pulse before the application of the CPMG sequence. Readout of the population along the x-axis $P_x$ is done with an ending $\pi/2$ and green laser pulse. (b) CPMGXY dynamical decoupling spectroscopy signal with $N=32$. Dips can be identified corresponding to entanglement with $^{13}\mathrm{C}$ nuclear spins. (b) Line fit (SI eq. 3) of individual dip centroids following algorithmic decomposition. Three spins were identified with parallel components: $A_1 = -20.34\pm0.70$ kHz, $A_2=35.03 \pm1.08$ kHz, $A_3=55.18\pm1.71$ kHz. (c) Fit of individual resonances to the analytical function (SI eq. 2) yielding parallel (A) and perpendicular (B) hyperfine components for each fitted $^{13}\mathrm{C}$ nuclear spin i: $(A_1,B_1) =-18.64 \pm 0.18, 18.36 \pm 0.69$ kHz, $(A_2,B_2) = 32.74 \pm 0.15, 28.76 \pm 1.08$ kHz, $(A_3,B_3) = 53.27 \pm 0.27, 34.37 \pm 1.65$ kHz. Uncertainties of each fitted A and B value represent $\pm$ 1 standard deviation.}
    \label{fig:3}
\end{figure*}

We perform CPMG dynamical decoupling measurements and observe sharp, periodic resonances indicative of coherent coupling to individual $^{13}\mathrm{C}$ nuclear spins (Fig. \ref{fig:3}a). In addition, a broader set of background resonances arise from overlapping contributions of the surrounding spin bath, while the overall signal envelope decays due to electron spin decoherence governed by $T_1$ and $T_2$. To extract hyperfine parameters, we apply an algorithmic decomposition procedure \cite{OhEtAl2020} to identify dip positions associated with individual nuclear spins. The dip centers corresponding to a given spin are fit to a linear relation (SI eq.~3), yielding an estimate of the parallel hyperfine coupling $A$ (Fig. \ref{fig:3}b). This estimate serves as an initial parameter for fitting the analytical model of the NV electron spin population $P_x$ (SI eq.~2) to individual resonances, enabling extraction of the perpendicular hyperfine component $B$ (Fig. \ref{fig:3}c).

Using this approach, we resolve three distinct $^{13}\mathrm{C}$ nuclear spins with their corresponding parallel (A) and perpendicular (B) hyperfine components at room temperature using only electron spin control. The extracted hyperfine parameters are $(A_1, B_1) = (-18.64 \pm 0.18, 18.36 \pm 0.69)$~kHz, $(A_2, B_2) = (32.74 \pm 0.15, 28.76 \pm 1.08)$~kHz, and $(A_3, B_3) = (53.27 \pm 0.27, 34.37 \pm 1.65)$~kHz. These results demonstrate that the QICK-DAWG platform provides all essential functionality required for CPMG-based nuclear register characterization.

Further identification of additional $^{13}\mathrm{C}$ spins is limited in this device by the laser-induced reduction of $T_1$, associated with the use of a single-pass AOM for optical control. To highlight the advantages of the improved timing resolution afforded by our platform, we extend our measurements to a second NV center at low temperature, employing a double-pass AOM configuration.

\subsection{High-Resolution Spectroscopy at Low Temperature}
\begin{figure*}[t]
    \centering
    \includegraphics[width=1\textwidth]{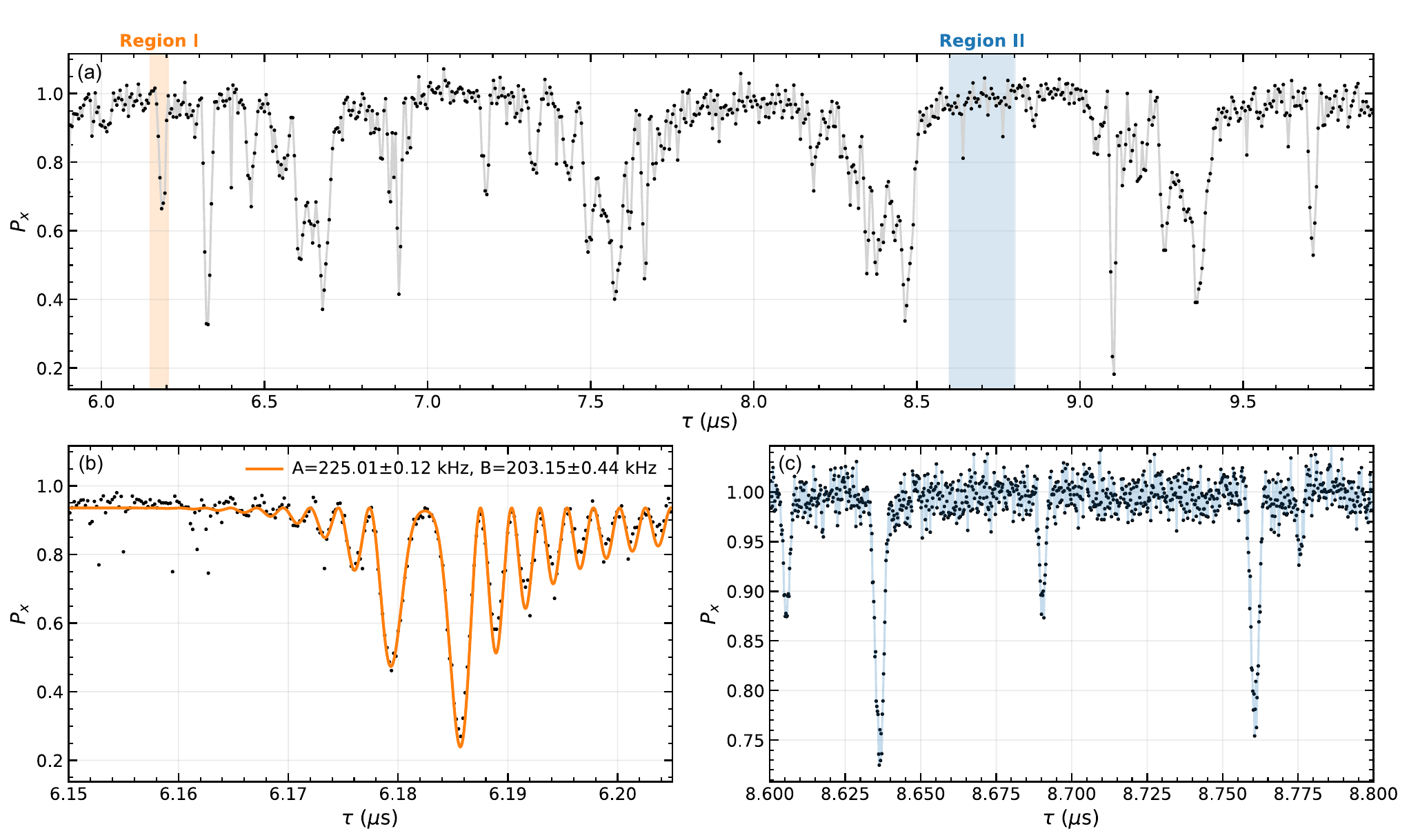}
    \caption{Higher resolution CPMG nuclear spin spectroscopy on a different NV from Fig. 3 at 4K. All MWs were performed with QICK. Lasers were controlled with ARTIQ. Magnetic field was at 525 Gauss. The CPMGXY pulse sequence was used with starting and ending $\pi/2$ pulses. (a) Low resolution, sampled with 5.09~ns timesteps, $N=32$ scan with off-resonant 532~nm green readout showing isolated dips from individual nuclear spins. Highlighted regions I and II were further run with 203 ps timesteps in the lower panels. (b) Region 1 consists of increasing $N=320$ to capture the sinusoidal oscillation produced by a single peak and fitted to SI Eq. 2 to extract the hyperfine components. (c) Region 2 at $N=32$ features multiple narrow peaks with resolvable widths at higher timing resolution. Bottom two scans were done with resonant readout and optical pumping \cite{robledoHighfidelityProjectiveReadout2011}.}
    \label{fig:4DQP}
\end{figure*}

The sensitivity of CPMG-based spectroscopy can be improved by increasing the number of dynamical decoupling pulses or extending $\tau$ to longer times, thereby enhancing spectral selectivity and isolating more nuclear spins from the bath \cite{TaminiauEtAl2012}. We therefore use a complementary mode of operation using a separate confocal setup at 4~K under a 525 G aligned magnetic field. In this setup, ARTIQ \cite{kasprowiczARTIQSinaraOpen2020} is the main optical and readout control system while the same QICK-based MW framework is retained \cite{RiendeauEtAl2023}, now leveraging its 200~ps timing resolution to access finer features in the dynamical decoupling signal. In this low-temperature environment, longer coherence times ($T_2=1.05$~ms) and resonant optical control \cite{robledoHighfidelityProjectiveReadout2011} with a double pass AOM system for improved extinction enable clearer resolution of individual resonances following application of the CPMG sequence (Fig. \ref{fig:4DQP}a). 

At low pulse numbers (e.g., $N=32$), the CPMG response exhibits a single broad resonance dip, which can lead to difficulty in fitting from local minima arising from many distinct combinations of hyperfine parameters (region I in Fig. \ref{fig:4DQP}a). As the number of pulses is increased to $N=320$, this broad feature evolves into a series of fine oscillatory fringes (Fig. \ref{fig:4DQP}b). These oscillations originate from a sinusoidal modulation superimposed on the intrinsic Lorentzian lineshape of the $^{13}\mathrm{C}$ spin signal (SI eq.~2). This structure at higher $N$ strongly constrains the allowed parameter space which is in contrast to the degenerate solutions obtained from a single broad dip at low $N$, enabling sub-kHz determination of the hyperfine couplings. Leveraging the $200$~ps timing resolution of our QICK-DAWG implementation, we densely sample these fringes and directly fit the analytical model across multiple oscillations with high precision yielding $(A, B) = (225.01 \pm 0.12, 203.15 \pm 0.44)$~kHz, thereby overcoming under-sampling limitations. This approach enables accurate extraction of hyperfine parameters from a single high-$N$ resonance, showing an alternative method to fitting multiple dips across different oscillation periods.

Finally, at high timing resolution with $N=32$, we observe additional narrow resonance dips that are not resolvable in lower-resolution measurements (Fig. \ref{fig:4DQP}c). These features cannot be accurately described by a single spin analytical model (SI eq.~2) under physically reasonable bounds on the $^{13}\mathrm{C}$  hyperfine parameters ($|A|, |B| < 1$~MHz).

Several mechanisms may give rise to such narrow features in the CPMG response. One possibility is coupling to spins with a larger effective gyromagnetic ratio, which would modify the resonance condition and produce sharper spectral features. Alternatively, strongly coupled nuclei with hyperfine interactions exceeding the assumed bounds could lead to narrow resonances in $\tau$-space. 

In all cases, resolving such features requires fine temporal sampling. Ongoing work is focused on identifying their physical origin through extended modeling and complementary measurements. These observations highlight the critical role of high timing resolution enabled by our QICK-DAWG platform for uncovering previously inaccessible structure in dynamical decoupling spectroscopy, thereby enhancing nuclear spin detection and control.

\section{conclusion}
In this work, we demonstrate sub-nanosecond timing control for spin-defect quantum memories using an inexpensive FPGA-based platform, achieving an effective resolution of 200~ps through a waveform-offset method on an RFSoC device. This approach overcomes the native $\sim$3.25~ns timing granularity of the FPGA sequencer without hardware modification, enabling high-resolution control within a compact and flexible architecture.

We apply this capability to dynamical decoupling spectroscopy of nitrogen-vacancy centers, demonstrating both rapid room-temperature characterization and high-resolution measurements at low temperature. The improved timing resolution enables dense sampling of narrow spectral features, allowing detailed lineshape fitting and extraction of hyperfine coupling parameters with sub-kHz precision. These results show that precise, device-level characterization of nuclear spin environments around spin defects can be achieved without reliance on expensive AWG-based systems.

Improved timing resolution directly impacts the scalability of spin-based quantum memories by expanding the set of resolvable and controllable nuclear spins. The integrated, programmable nature of the QICK platform further enables experiment-in-the-loop approaches, providing a pathway toward automated, AI-assisted characterization and optimization of color center based quantum memories \cite{varona-uriarteAutomaticDetectionNuclear2024}.

\section*{Acknowledgment}

We thank Sho Uemura and Mo Chen for helpful discussions; Toby Chu for wirebonding assistance.  This work was performed, in part, at the Center for Integrated Nanotechnologies, an Office of Science User Facility operated for the U.S. Department of Energy (DOE) Office of Science. Sandia National Laboratories is a multimission laboratory managed and operated by National Technology \& Engineering Solutions of Sandia, LLC, for the U.S. DOE’s National Nuclear Security Administration under contract DE-NA-0003525. This work was supported by Sandia National Laboratories’ Laboratory Directed Research and Development (LDRD) program.  Part of this work was conducted at the Washington Nanofabrication Facility, a National Nanotechnology Coordinated Infrastructure (NNCI) site at the University of Washington, supported in part by the National Science Foundation (NNCI-1542101, NNCI-2025489).  This work was supported by NIST (60NANB23D202), NSF PHY-GRS 2233120, the National Science Foundation (NSF) Center for Integration of Modern Optoelectronic Materials on Demand (IMOD) under Grant DMR-2019444, the Quantum Leap Challenge Institutes (Quantum Systems through Entangled Science and Engineering) under Grant OMA-2106244, and by the U.S. Department of Energy, Office of Science, National Quantum Information Science Research Centers, Co-design Center for Quantum Advantage (C2QA) under contract number DE-SC0012704, (Basic Energy Sciences, PNNL FWP 76274)

\section{Supplemental Information}
The Hamiltonian of an NV spin coupled with a single nuclear spin in the rotating frame and in the secular approximation is given by:
\begin{equation}
    \hat{H} = \ket{0}\bra{0}\hat H_0 + \ket{1}\bra{1}\hat H_1
\end{equation}
where $\hat H_0 =\omega_L\hat I_z $ and $\hat H_1 =(A+\omega_L)\hat I_z +B\hat I_x$. $\omega_L$ is the nuclear Larmor frequency, A and B are the parallel and perpendicular component of the hyperfine coupling respectively. For a single nuclear spin, the probability $P_x$ of preserving the initial electron spin state under the CPMG sequence is given by: 
\begin{equation}
    P_x = \frac{M+1}{2}, \quad M = 1-(1-\hat{n}_{0} \cdot \hat{n}_{1})\sin^2(\frac{N\phi}{2})
\end{equation}
where N is the number of $\tau-\pi-\tau$ units in the CPMG sequence, $\hat n_0 $ and $ \hat n_1 $ are rotation axes of the nuclear spin dependent on the initial electron spin state, and $\phi$ is the rotation angle around such an axis. Under a large magnetic field ($\omega_L \gg B$), the position of the kth resonance is given by
\begin{equation}
    \tau_k = \frac{(2k-1)\pi}{A+2\omega_L}
\end{equation}
Detailed derivations and expressions can be found in \cite{TaminiauEtAl2012}\cite{OhEtAl2020}.

\bibliographystyle{IEEEtran}
\bibliography{2026_QCE_Paper}
\end{document}